\documentstyle[preprint,prl,aps,epsfig]{revtex}
\def\DESepsf(#1 width #2){\epsfxsize=#2 \epsfbox{#1}}
\def\CP{$CP$}
\begin{document}
\pagestyle{empty}                                      
\preprint{
\hbox to \hsize{
\hbox{
            }
\hfill $
\vtop{
 \hbox{NTUHEP-99-25}
 \hbox{COLO-HEP-438}
 \hbox{LNS-99-290}
 \hbox{ }}$
}
}
\draft
\vfill
\title{Determination of the Phase of \boldmath{$V_{ub}$} 
\\ from Charmless Hadronic $B$ Decay Rates
}
\vfill
\author{$^{1}$Wei-Shu Hou, $^{2}$James G. Smith
 and $^{3,4}$Frank W\" urthwein}
\address{
\rm $^1$Department of Physics, National Taiwan University,
Taipei, Taiwan 10764, R.O.C.\\
\rm $^2$Department of Physics, University of Colorado,
Boulder, CO 80309, USA\\
\rm $^3$Lauritsen Lab, California Institute of Technology,
Pasadena, CA 91125, USA\\
\rm $^4$Laboratory of Nuclear Studies,
Massachusetts Institute of Technology, Cambridge,~MA~02139,~USA
}

%
%
\vfill
\maketitle
\begin{abstract}
We perform a model dependent fit to recent data on charmless
hadronic $B$ decays and determine $\gamma$, 
the phase of $V_{ub}^*$. 
We find $\gamma = 114^{+25}_{-21}$ degrees,
which disfavors the often quoted $\gamma \sim 60^\circ$ at the two standard 
deviation level.
We also fit for the form factors $F_0^{B\pi}$ and $A_0^{B\rho}$,
and the strange-quark mass.
They are consistent with theoretical expectations,
although $m_s$ is somewhat low.
Such agreement and the good $\chi^2$ for the fit
may be interpreted as a 
confirmation of the adequacy of our model assumptions.
\end{abstract}
\pacs{PACS numbers: 
}

\pagestyle{plain}

The measurement of a surprisingly large 
$\varepsilon^\prime/\varepsilon$ value in 1999
is an exasperating reminder of 
how little we really know about \CP\ violation in Nature.
Within the Standard Model (SM) with 3 quark generations,
however, there is a unique phase \cite{KM} in the
Kobayashi-Maskawa (KM) matrix $V$,
often defined as $\gamma = {\rm arg}(V_{ub}^*)$
in the usual phase convention \cite{PDG}.
At present, there is no evidence that
this phase fails to account for \CP\ violation phenomena.

Two $B$ factories, built to study \CP\ violation in the $B$ system, 
have just been completed. 
By comparing the time dependence of tagged 
$B^0$ vs. $\overline B^0 \to J/\psi K_S$ decays, 
one can cleanly measure 
the \CP\ phase in $B^0$-$\overline B^0$ mixing,
which, in the SM, gives $\sin2\beta$ where $\beta = {\rm arg}(V_{td}^*)$.
Together with the demonstrated capabilities of 
collider detectors at the Tevatron, 
a precise measurement of $\sin2\beta$ is assured within a year or two.
The unitarity phase $\alpha$ can be measured 
via $\pi^+\pi^-$ or $\pi^+\pi^-\pi^0$ modes 
but now appears to be more challenging because
the $\pi^+\pi^-$ rate is smaller than expected,
which in turn implies larger ``penguin pollution". 
However, it is the phase $\gamma$ that is 
usually viewed as the most difficult.
All suggestions so far require very high statistics
or various technical challenges. 
In this Letter we exploit the emerging rare $B$ decay data
from CLEO and perform a fit~\cite{pipiLepPho} that,
though model dependent, allows extraction of
$\gamma$ with just $10^7$ $B$ mesons
using only \CP-averaged rates.
The goodness of fit and reasonableness of other fit parameters
serve as checks on the adequacy of our model assumptions.
Since neither vertexing nor tagging is required,
this method will benefit from the improved statistics soon available from
the CLEO upgrade as well.

The measurement of $\sin2\beta$ is often
compared to a double-slit interference 
experiment,
the two slits being $B^0$ and $\overline B^0 \to J/\psi K_S$ decays.
%
Charmless rare $B$ decay rates,
even when CP averaged, can also be viewed as double-slit experiments
that in principle probe the phase~$\gamma$.
%
The present observed pattern
that $\overline B\to \overline K\pi$ rates are larger than $\pi\pi$
but comparable to $\rho^0\pi^-$, $\rho^\mp\pi^\pm$ and $\omega\pi^-$
%
%
implies that 
both tree (T) and penguin (P) amplitudes contribute
to these rates, hence the double-slit analogy.
Unfortunately, hadronization effects such as 
final state interactions (FSI) could dilute such interference effects.

The Fleischer-Mannel bound \cite{FM}
on $\gamma$ is no longer effective since one now has 
$R \equiv \overline\Gamma(B^-\to K^-\pi^+)/
     \overline\Gamma(\overline B^0\to K^0\pi^+) = 1.0 \pm 0.3$,
where $\overline\Gamma$ denotes 
the average of $B$ and $\overline B$ widths. 
%
A more promising method \cite{NR} is based on 
$R_* = \overline\Gamma(B^-\to \overline K^0\pi^-)/
      2\overline\Gamma(B^-\to K^-\pi^0)$,
with some reference to $\pi\pi$ for control of model dependence.
But with $R_* = 0.75 \pm 0.30$ at present,  
one cannot set a useful bound on $\gamma$.  More than an order of 
magnitude increase in data is needed for a restrictive measurement. 
In this Letter we take a more global view and perform fits which
trade model independence for exhaustive use of available data.
We also use only \CP-averaged rates, since 
there is no sign of significant \CP\ asymmetries \cite{CPALepPho} 
and the errors are large. 
%
Asymmetries in addition are more sensitive to FSI than
averaged rates.

We shall assume that naive factorization ($N_c = 3$) is a good approximation,
and use effective-theory matrix elements cross-checked 
by two groups \cite{ali,CCTY}, ignoring annihilation type diagrams.
We make a $\chi^2$ fit of data to $\gamma$ and four other parameters.
Factorization in two body rare $B$ decays
may be heuristically justified by the large energy release \cite{Bj}:
final state mesons move away from each other
so fast that they do not interact.
Recent theoretical work suggests that factorization 
may be derivable from QCD in certain limits~\cite{theoryfact}.
In our view, factorization provides 
a simple framework to describe hadronic $B$ decays that is 
rich in predictions with a limited set of free parameters.
It is therefore reasonable to use this framework when attempting a
first global fit to the large number of results on charmless hadronic
$B$ decays now available. 
This work is motivated by Ref. \cite{HHY},
which pointed out that
recent CLEO rare $B$ data supports factorization 
if $\cos\gamma < 0$ is taken.
Indirect fits  
to the unitarity triangle find a 95\% C.L. range for $\gamma$
of $44^\circ$ -- $75^\circ $~\cite{parodi}, $44^\circ$ -- $93^\circ$
\cite{mele}, $41^\circ$ -- $97^\circ $~\cite{plaszczynski},
and $36^\circ$ -- $97^\circ $~\cite{ali-london}, 
depending in part on how conservatively the theoretical errors are
handled.

Let us illustrate the parameters that enter with 
$\overline B^0\to K^-\pi^+$,
which is a $b\to s\bar uu$ transition under factorization. 
Ignoring annihilation terms, one has \cite{ali,CCTY}
\begin{eqnarray}
 {\cal A}_{K^-\pi^+} \propto
 f_K F_0^{B\pi } \; (m_B^2-m_\pi^2)
 \left\{V_{us}^*V_{ub} a_1 - V_{ts}^*V_{tb} \left[a_4+a_{10}  
                       +(a_6+a_8)R_{su}\right]\right\}.
\label{eq:kpi}
\end{eqnarray}
%
We are free to fix $\vert V_{ts}\vert \cong \vert V_{cb}\vert = 0.0381$ 
\cite{PDG} since any uncertainty can be absorbed in form factors.
The two relevant fundamental parameters are therefore
$\vert V_{ub}/V_{cb}\vert$ and $\gamma$, and the latter clearly 
controls the interference between tree and penguin terms.
The parameters $a_i$
are related to short distance Wilson coefficients (WC)
and evaluated within a QCD framework.
They also depend on the scale parameter $\mu_f$ 
where factorization is operative. 
The values of $a_i$ in the literature are still evolving as
issues of scale, scheme and gauge dependence are addressed.
We use two sets of $a_i$ from Refs.~\cite{ali} (AKL) and~\cite{CCTY} (CCTY).  
The dominant strong penguin coefficients
are $a_4$ and $a_6$ ($-0.04$ to $-0.06$),
while the dominant electroweak penguin coefficient is 
$a_9 \simeq -0.009$ coming from the $Z$ penguin.
We use the $a_i$ for $b\to s$ since the difference for $b\to d$ is small.

In Eq. (\ref{eq:kpi}) one also has the factor
\begin{eqnarray}
  R_{su} = 2m_{K^-}^2/(m_b-m_{u})(m_s+m_u) ~.
\label{eq:rsu}
\end{eqnarray}
%
A similar factor $R_{sd}$, which enters ${\cal A}_{\bar K^0\pi^-}$,
is taken to be equal to $R_{su}$.
This factor can be better understood as a product of two pieces:
the factor $1/(m_b - m_u) \cong 1/m_b$  
balances against $m_B^2 - m_\pi^2$;
and $m_{K^-}^2/(m_s+m_u)$ is nothing but the nonperturbative part
of the pseudo-Goldstone boson mass formula, which is 
well defined within QCD but not yet very well determined.
Although $R_{su}$ is technically related to an $m_s$-independent
hadronic matrix element,
in the form of Eq. (\ref{eq:rsu}), it becomes a sensitive
probe of $m_s$ in a way that is 
analogous to $K\to \pi\pi$ decay and $\varepsilon^\prime/\varepsilon$. 

The factors $f_K$ and $F_0^{B\pi}$ 
arise from evaluating hadronic matrix elements of 
four quark operators under factorization:
The former comes from forming $K^-$ out of the vacuum 
via the $\bar s\gamma_\mu\gamma_5 u$ current,
the latter arises from the transition 
$\bar B^0\to\pi^+$ via the $\bar u\gamma_\mu b$ current.
While form factors are well defined, 
it is the reliance on models
that causes us to lose track of the factorization scale $\mu_f$.
Popular form factor models are the BSW model 
and light-cone sum rule (LCSR) evaluations.
A recent compilation of models can be found in~\cite{deandrea},
but we shall treat form factors as fit parameters. 

The criteria for choosing the decay modes to include in the fit
are as follows.  
First, 
a central value branching ratio (BR) 
with statistical and systematic errors must be available.  
Second, 
we exclude $VV$ modes such as $\omega K^*$
since there is insufficient data 
to constrain the extra form factors that enter.  
Third,
we require that the experimental sensitivity
(a few times $10^{-6}$ at present) to be 
comparable to the range of factorization predictions.
%
Only the $\omega\pi^0$ final state,
with a predicted BR below $10^{-7}$, 
is removed with this criterion.
Since this and other suppressed decays such as $\rho^0\pi^0$, $\pi^0\pi^0$, 
$\phi\pi$, $K\overline K$ and $K^*\overline K$
may well be dominated by FSI from other charmless final states, 
factorization is less likely to work well. 
We therefore propose to exclude these modes 
even when suitable measurements become available.
The only exceptions to these rules are 
final states involving $\eta$ and $\eta^\prime$. 
We prefer to apply the fit to predict their BRs 
rather than use them in the fit 
because the $q\bar q$ content and other issues of 
these mesons have recently been questioned.
We note that the newly measured \cite{etaLepPho} 
$\eta K^*$ modes, like $\eta^\prime K$, 
are larger than previous theoretical predictions~\cite{ali,CCTY}. 

We give the 14 measured BRs (averaged over $B$ and $\overline B$)
that enter our fit in Table I,
where we also give the fitted output.  
To limit the number of fit parameters,
we use approximate relations as follows.
We assume KM unitarity hence
$-V_{ts}^*V_{tb} \cong \vert V_{cb}\vert$ and
$-V_{td}^*V_{tb} = V_{cd}^*V_{cb} + V_{ud}^*V_{ub}
             \cong - \vert V_{cb}\vert (\lambda 
                 - e^{-i\gamma} \vert V_{ub}/V_{cb}\vert)$,
where $\lambda = \vert V_{us}\vert$.
Since $\lambda - \vert V_{ub}/V_{cb}\vert\cos\gamma >0$ always,
as noticed in Ref. \cite{HHY}, 
T-P interference is
opposite in sign for P-dominated and T-dominated modes
such as $K^-\pi^+$ and $\pi^-\pi^+$, 
leading to enhanced $K^-\pi^{+,0}$ and suppressed $\pi^-\pi^+$
for $\cos\gamma < 0$, in better agreement with data.
The chiral relation $m_{K^-}^2/m_{\pi^-}^2 \simeq (m_s+m_u)/(m_d+m_u)$
and the fact that $m_s \gg m_{d,u}$
give $R_{su} \cong R_{sd} \cong R_{du}$,
while $Q_{ij} = - R_{ij}$ for $VP$ modes 
such as $\rho\pi$, $\omega\pi$ and $\omega K$.
We use form factors at $q^2 = 0$ and
$F_1^{BP} = F_0^{BP}$; 
$F_0^{BK}/F_0^{B\pi} = 1.13$ which is
consistent with both BSW and LCSR models;
$A_0^{B\omega} = A_0^{B\rho}$; 
and $A_0^{BK^*} = 1.26\, A_0^{B\rho}$ (used for predictions only). 
Surveying the amplitudes for modes in Table I,
we find that just five parameters suffice for the fit:
$\gamma$, $\vert V_{ub}/V_{cb}\vert$, $R_{su}$,
$F_0^{B\pi}$ and $A_0^{B\rho}$.

The function minimized by the fit is 
\begin{equation}
\chi^2=\sum_i 
   ( ({\rm BR}^i_{\rm meas} - {\rm BR}^i_{\rm pred})/\sigma^i_{\rm meas})^2
	+ ((0.08 - |V_{ub}/V_{cb}|_{\mathrm{pred}})/0.02)^2~,
\label{eq:chisquare}
\end{equation}
where we sum over the modes in Table I.  
The predicted BRs are calculated from formulas like 
Eq. (\ref{eq:kpi}) taken from Refs. \cite{ali,CCTY}.
We have checked that we confirm the $N_c = 3$
results of AKL and CCTY with the same input parameters.
We take into account the full (asymmetric) experimental errors
and correlations in $K^-\pi^+/\pi^-\pi^+$, $K^-\pi^0/\pi^-\pi^0$ 
and $\omega K^-/\omega\pi^-$ measurements, 
where the correlation coefficients 
are $-0.15$, $-0.29$ and $-0.17$, respectively.
The fit is able to nearly optimally use the information for each
of these modes individually, though $K/\pi$ separation improvements
in the next round of experiments will help in this regard.
For simplicity, we assume that systematic errors 
have the same correlation coefficient as statistical errors,
i.e. we apply the correlation coefficient to the total error
with all errors combined in quadrature. 
If the best fit value is below the experimental central value, 
the low-side experimental error is used, 
and conversely the high-side.

To understand the behavior of the $\chi^2$ function 
in the 5D fit parameter space,
its dependence on various fixed parameters
or the exclusion of certain experimental measurements,
we have explored many variants of our nominal fit.  
%
In all cases we find $\gamma > 100^\circ$.  
%
Our nominal fit results, with CCTY $a_i$ values, are given in Table II.
The $\chi^2$ per degree of freedom (DOF) in the last column indicates 
the good quality of the fit.  We choose CCTY
rather than AKL as nominal only because of their claim of improved gauge
dependence of the $a_i$.  The fit values for AKL (see Table II) differ
only slightly from our nominal, and mostly because of the larger $\vert
a_{4,6}\vert$ found by these authors.
  We note that $R_{su}$ for AKL input should be smaller than
  the CCTY case since quark masses are defined at $\mu = 2.5$~GeV
  rather than $m_b$.
  We have also checked that the strong phases of $a_{4,6}$
  have little impact on our fitted $\gamma$ value.

\begin{figure}[htb]
\vspace{0.5cm}
\centerline{{\epsfxsize3.2 in \epsffile{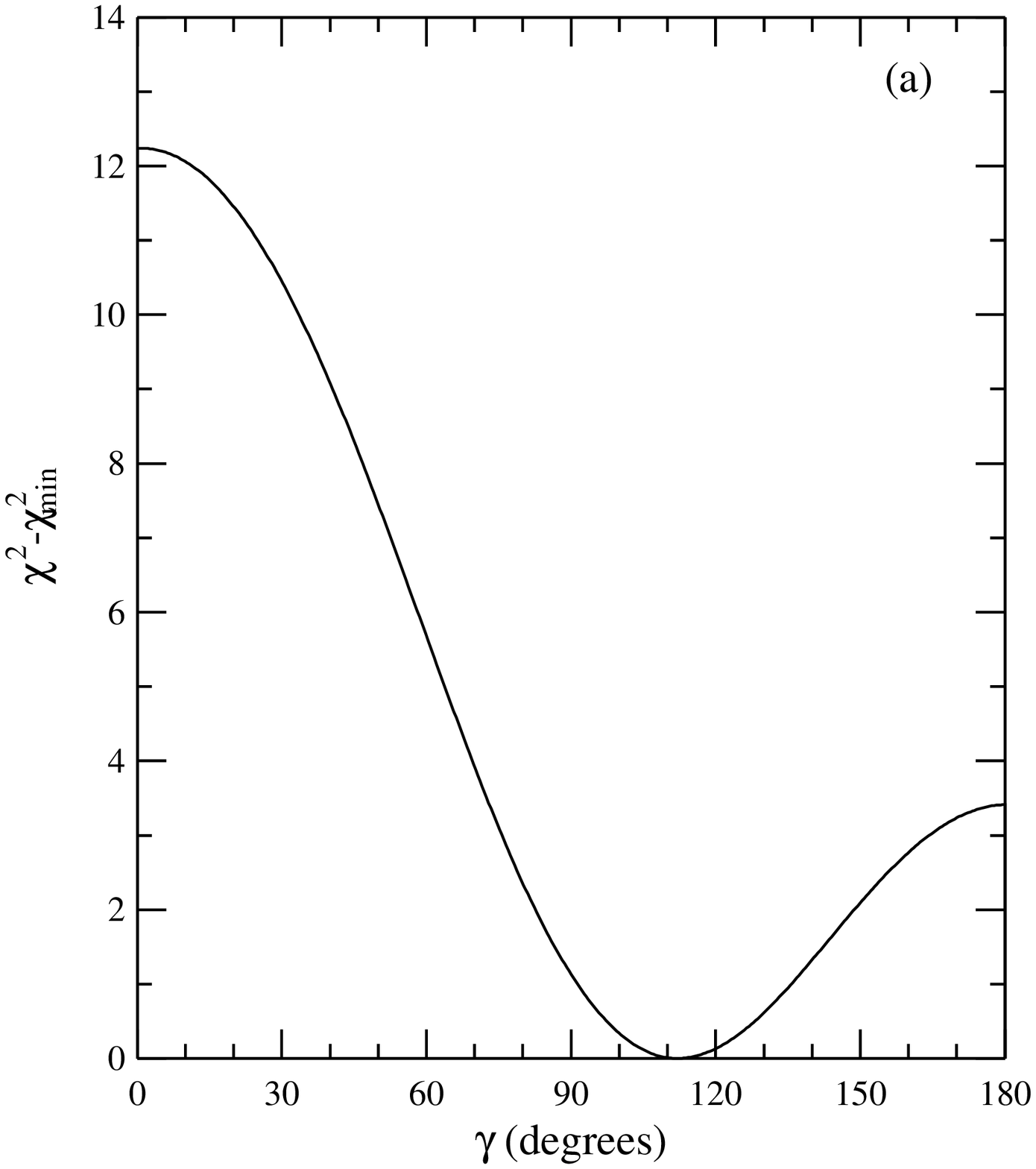}}
            {\epsfxsize3.2 in \epsffile{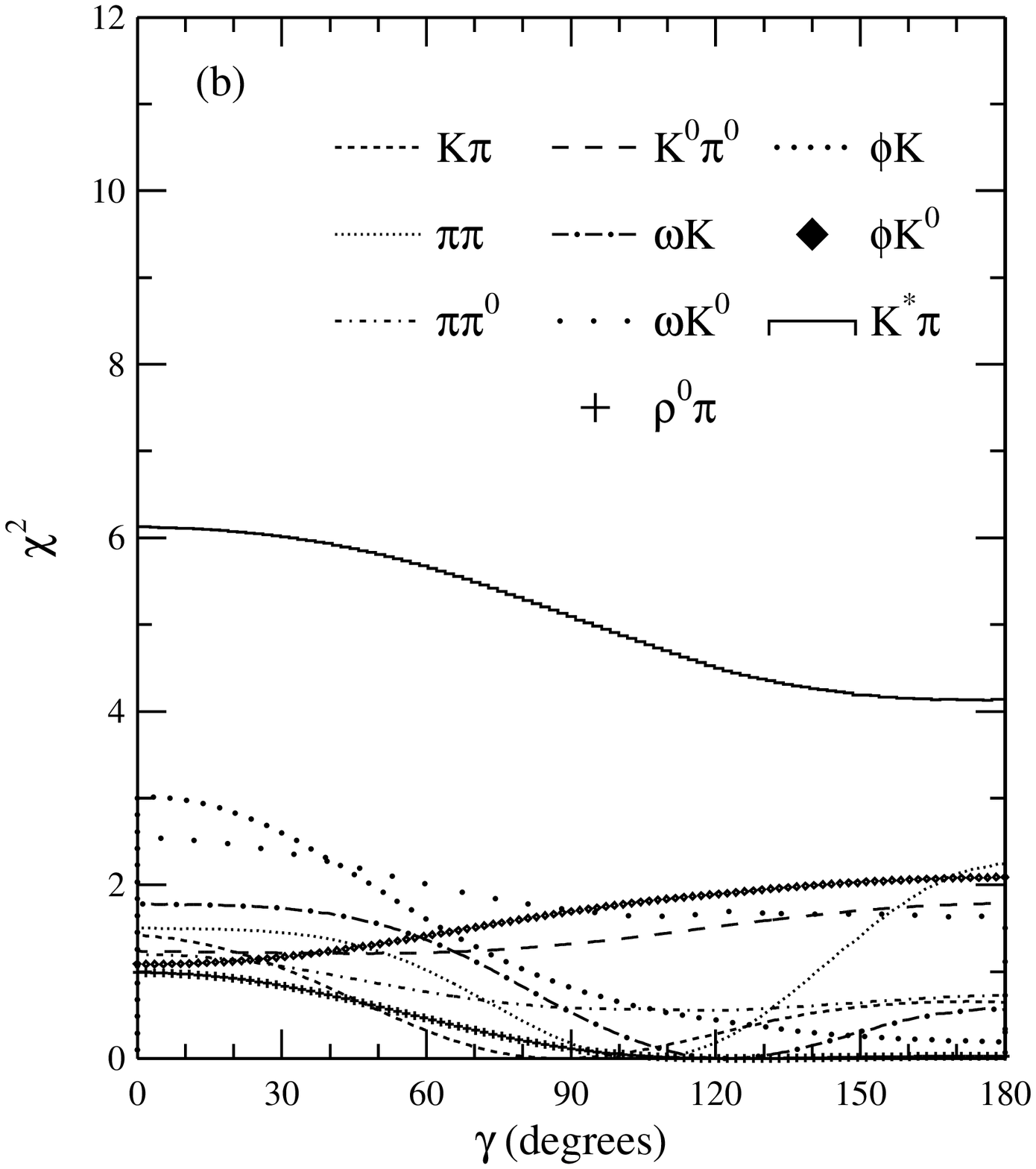}}}
\vspace{0.5cm}
\caption{
(a) $\chi^2 - \chi^2_{\rm min}$ vs. $\gamma$ from nominal fit,
    where $\chi^2_{\rm min} = 11.3$ for 10 DOF;
(b) Major $\chi^2$ contributions from individual measurements.
}
\label{fig:gamma}
\end{figure}

The $\chi^2$ vs. $\gamma$ curves for the nominal fit with CCTY input
are shown in Fig. \ref{fig:gamma}. 
We note that our $\gamma$ value has a two-fold ambiguity since
the fit is sensitive to $\cos\gamma$ rather than $\gamma$.
From the contributions from individual modes
given in Fig. \ref{fig:gamma}(b),
we see that the main discriminator 
for favoring large $\gamma$ comes from $K^-\pi$ and $\pi^-\pi$ modes,
and, somewhat surprisingly, the $\omega K^-$ and $\phi K^-$ modes.
The situation for $\phi K^-$ is a result of the procedure of
minimizing $\chi^2$ for each $\gamma$ value.
This induces an apparent sensitivity, due to changes in the other
parameters, where there is no direct dependence.

  The error on $|V_{ub}/V_{cb}|$ returned by the fit is
  only marginally better than the conservative range 
  $\vert V_{ub}/V_{cb}\vert =0.08 \pm 0.02$ \cite{PDG}
  used as an additional term in Eq. (\ref{eq:chisquare}).
  Sensitivity to $|V_{ub}/V_{cb}|$ largely comes from
  $\overline B \to \rho^0\pi^-,\ \rho^\pm\pi^\mp$ and
  $\omega\pi^-$, all of which depend on $A_0^{B\rho}$. 
  Removing the constraint on $|V_{ub}/V_{cb}|$ from the fit
  gives higher $|V_{ub}/V_{cb}|$ with large errors and
  strongly correlated with $A_0^{B\rho}$
  (and between $R_{su}$ and $F_0^{B\pi}$ as well) but 
  $\gamma = 121^{+31}_{-24}$ degrees remains close to nominal
  (see Table II).

The fit favors large $R_{su}$ ($R_{du}$)
since it facilitates the enhancement (suppression)
of $K^-\pi^{+,0}$ ($\pi^-\pi^+$) modes.
Furthermore, under factorization, the BRs for $\omega \overline K$ modes 
are enhanced only for large $R_{su}$ such that $a_4$ and $a_6$ 
penguin contributions do not cancel fully.  We have checked that when
the $\omega \overline K^0$ mode is removed from the fit, there is no
significant change in $\gamma$, though $R_{su}$ drops to 1.69.
Our nominal $R_{su}$ fit value implies 
$m_s  = 58^{+14}_{-11}$, $67^{+16}_{-13}$ MeV 
at $m_b$ and 2 GeV scale, respectively.
This is lower than what is 
commonly used in most previous calculations,
but consistent with recent unquenched lattice results 
which give $m_s(2\ {\rm GeV})=84 \pm 7$ MeV \cite{aoki}.
In addition recent experimental results for $\varepsilon^\prime/\varepsilon$
can be reconciled better with theoretical predictions
if a smaller value of $m_s$ is used \cite{aoki}.
For comparison, the result for fixing $R_{su}= 1.21$ ($m_s(m_b)=90$ MeV)
is given in Table II.
Note that $R_{su}$ is anti-correlated with $a_6$ 
since only the product appears in the amplitudes.
As for the form factors, our fitted $F_0^{B\pi}$ ($A_0^{B\rho}$)
is lower (higher) than but consistent with the LCSR result of 
$0.305\pm 0.046$ ($0.372 \pm 0.074$).

Predictions from our fit for some selected modes are given in Table III.
The agreement with the newly measured~\cite{etaLepPho} 
$\eta \overline K^*$ modes are rather striking.
An enhancement factor of 1.7 comes from $A_0^{BK^*}\simeq 0.60$ compared to 
LCSR value of $A_0^{BK^*} \sim 0.47$, the rest coming from our low $m_s$.
The $\eta^\prime \overline K^*$ modes are
comparable in size to the observed $\eta \overline K^*$ modes.
Since we can account for less than half the rate
of $\eta^\prime \overline K$ modes and the missing contribution 
may well be specific to the $\eta^\prime$ decay modes,
our predicted $\eta^\prime \overline K^*$ rates 
should be viewed with some caution.
The $\rho^-\pi^0$ mode is suppressed by $\cos\gamma < 0$,
smaller $F_1^{B\pi}$
(which also suppresses $\overline K^*\pi$ modes) 
plus destructive interference between two terms because of low $m_s$.
The $\rho K$ modes are enhanced by the low value of $m_s$
and larger $A_0^{B\rho}$ form factor, 
except for $\rho^0 K^-$, which is suppressed compared to 
$\rho^+ K^-$ by destructive interference
between the strong $a_6$ and electroweak $a_9$ penguin terms
(a similar effect suppresses $\overline K^{(*)0}\pi^0$ modes).

%
%

As an aside, we give the ``penguin pollution'' 
as determined from our fit.
Defining $T$ ($P$) as the amplitude arising from
$a_{\rm 1-2}$ ($a_{\rm 3-10}$),
we find the ratio $|P/T|$ in
$\pi^+\pi^-$ ($\rho^0\pi^\pm$) to be $0.37\pm 0.04$ ($0.20 \pm 0.04$)
for our nominal fit.
For comparison, the CCTY result for
$N_c = 3$, $\gamma \simeq 65^\circ$, $\vert V_{ub}/V_{cb}\vert = 0.090$,
$m_s(m_b) = 90$ MeV using LCSR form factors gives 0.20 (0.10).

How do we reconcile with the usual fit to $B$ and $K$ data
other than charmless rare $B$ decays, which give a 95\% C.L. 
range that excludes $\cos\gamma<0$ 
\cite{parodi}?
The removal of the second quadrant in these fits results mostly
from combining recent bounds on $B_s$ mixing from LEP, CDF and SLD,
with lattice QCD results that 
relate $B_d$ and $B_s$ mixing parameters \cite{mele}.
Thus, our results
suggest that $B_s$ mixing could be very close to the present limit.

It should be stressed that the goodness of our fit (see Table II)
suggests that corrections to factorization may be small compared 
with the present experimental precision.
It is reassuring that the hadronic parameters
from our fit are not at variance with theoretical expectations.
We note  that $\gamma$ is the most stable parameter in the fit, 
with $\cos\gamma<0$ for all variations we have considered.
This is because $\gamma$ directly controls the ``double-slit" interference,
while other parameters enter indirectly.
We note that our larger value of 
$\gamma$ tends to reduce the value of $\sin2\beta$.

In conclusion,
we have made a model-dependent determination of 
$\gamma = 114^{+25}_{-21}$ degrees. 
It will be interesting to see if future, more precise measurements will
confirm this result and the predictions in our tables.


%
This work is supported in part by grants from the US DOE and
NSC of Taiwan, R.O.C.
We thank our CLEO colleagues for the excellent data.

\begin{table}[htb]
\vspace{0.1cm}
 {\small Table I. Measured BR (in $10^{-6}$ units) 
\cite{pipiLepPho,etaLepPho}
  entering the fit and output BR from our fit,
  and for comparison (in parenthesis) the $N_c = 3$ values
  from Ref. \cite{CCTY} for $\gamma = 65^\circ$ and LCSR form factors.} 
\begin{center}
\begin{tabular}{ c l c  c l c } 
 Mode \hskip0.2cm &  Measured BR &  BR from Fit \hskip0.6cm &  
 Mode \hskip0.2cm &  Measured BR &  BR from Fit  \\ 
\hline
 $K^-\pi^+$ &  $18.8^{+2.8}_{-2.6} \pm 0.7 \pm 1.1$ &
 $20.9$ (12.9)  &
 $\pi^-\pi^+$ & $\ \, 4.7^{+1.8}_{-1.5} \pm 0.5 \pm 0.3$ &
 $\ \, 4.4$ (10.7)
  \\
 $K^-\pi^0$ &  $12.1^{+3.0+1.9}_{-2.8-1.2} \pm 0.8$ &
 $11.8$ ($\ \, 8.8$)  &
 $\pi^-\pi^0$ & $\ \, 5.4^{+2.1}_{-2.0} \pm 1.5\pm 0.3$ &
 $\ \, 3.4$ ($\ \, 5.7$)
  \\
 $\overline K^0\pi^-$ &  $18.2^{+4.6}_{-4.0} \pm 0.9\pm 1.3$ &
 19.5 (15.7) &
 $\overline K^0\pi^0$ &  $14.8^{+5.9+2.4}_{-5.1-3.3} \pm 2.5$ & 
 $\ \, 8.0$ ($\ \, 5.6$)
  \\
\hline
 $\rho^0\pi^-$ &  $15 \pm 5 \pm 4$ &
 14.0 ($\ \, 8.1$)  &
 $\rho^{\mp}\pi^{\pm}$ &  $35^{+11}_{-10} \pm 5$ &
 39.0 (41.3)
  \\
 $\omega \pi^-$ & $11.3^{+3.3}_{-2.9}\pm 1.0$ & 
 11.5 ($\ \, 8.2$)  &
 $K^{*-}\pi^+$ &  $22^{+8+4}_{-6-5}$ & 
 $\ \, 5.4$ ($\ \, 4.3$)
  \\
 $\omega K^-$ & $\ \, 3.1^{+2.4}_{-1.9} \pm 0.8$ &
 $\ \, 3.7$ ($\ \, 1.4$)  &
 $\phi K^-$ & $\ \, 1.6^{+1.9}_{-1.2} \pm 0.2$ &
 $\ \, 3.0$ ($\ \, 5.0$)  \\
 $\omega\overline K^0$ & $10.0^{+5.4}_{-4.2} \pm 1.5$ & 
 $\ \, 4.3$ ($\ \, 0.4$)  &
 $\phi\overline K^0$ & $10.7^{+7.8}_{-5.7} \pm 1.1$ & 
$\ \, 2.8$ ($\ \, 4.6$)
    \\
\end{tabular}
\end{center}
\end{table}

\begin{table}[htb]
 {\small Table II. Results for fitted parameters.
  CCTY (AKL) implies use of $a_i$'s
  from Ref. \cite{CCTY} (Ref. \cite{ali})}.
\begin{center}
\begin{tabular}{ l c  c  c c c c } 
 Input $a_i$ &  $\gamma$ (degrees)  &  $|V_{ub}/V_{cb}|$  &  $R_{su}$  &  
    $F_0^{B\pi}$  &  $A_0^{B\rho}$  & $\chi^2_{\rm min}/{\rm DOF}$\\ 
\hline
CCTY  &  $114^{+25}_{-21}$  &  $0.087 \pm 0.016$  & $1.89^{+0.43}_{-0.36}$  &
   $0.26\pm 0.04$   &  $0.48^{+0.13}_{-0.10}$  & 11.3/10
  \\
AKL  &   $105^{+23}_{-21}$  &  $0.092 \pm 0.016$  &  $1.70^{+0.34}_{-0.28}$ &
   $0.23 \pm 0.03$  &  $0.45^{+0.12}_{-0.09}$  & 12.2/10
  \\
CCTY  &  $124^{+56}_{-29}$  &  $0.076 \pm 0.014$  & 
$1.21$ (fixed) & $0.32\pm 0.03$  & $0.58^{+0.15}_{-0.11}$  & 15.0/11
  \\
CCTY  &  $121^{+31}_{-24}$  &  $0.105^{+0.063}_{-0.032}$ (free) & 
$2.21^{+1.30}_{-0.61}$ & $0.23^{+0.06}_{-0.07}$&$0.48^{+0.13}_{-0.10}$ & 11.0/9
\end{tabular}
\end{center}
\end{table}

\begin{table}[htb]
 {\small Table III. Predicted BR (in $10^{-6}$ units) 
  with CCTY $a_i$s for some modes not used in the fit,
and (in parenthesis) from CCTY for
$N_c = 3$, $\gamma = 65^\circ$, $\vert V_{ub}/V_{cb}\vert = 0.090$,
$m_s(m_b) = 90$ MeV using LCSR form factors.
For $\eta^\prime$, $\eta$ modes,
we have used
$f^u_{\eta^\prime}$, $f^s_{\eta^\prime} = 64$, 141 MeV, 
$f^u_{\eta}$, $f^s_{\eta} = 78$, $-113$ MeV, and 
$F_0^{B\eta^\prime}$, $F_0^{B\eta}=
0.108$, $0.121$, respectively~\cite{ali,CCTY}.
The $f^c_{\eta^{(\prime)}}$ effects are small.}
\begin{center}
\begin{tabular}{ c l c | c c | c c | c c} 
  Mode  \hskip0.1cm &  Measured BR\cite{etaLepPho} \ \  &  BR$_{\rm pred}$  \ \  
  &  Mode  &  BR$_{\rm pred}$  \ \    &  Mode  &  BR$_{\rm pred}$  \ \  
  &  Mode  &  BR$_{\rm pred}$  \\ 
\hline
 $\eta^\prime K^-$ & $80^{+10}_{-9} \pm 8$ & 32.4 (21.9)  &
 $\eta K^-$ & $\ \, 1.9$ (1.7)    &
 $\rho^-\pi^0$  &  $\ \, 6.5$ (15.0)   &
 $\rho^+ K^-$   &  $\ \, 6.1$ (2.0)
  \\
 $\eta^\prime\overline K^0$ & $88^{+18}_{-16} \pm 9$ & 28.1 (21.7)  &
 $\eta \overline K^0$ & $\ \, 1.9$ (1.0)   &
 $K^{*-}\pi^0$  &  $\ \, 5.1$ ($\ \, 3.3$)   &
 $\rho^0 K^{-}$ &  $\ \, 1.8$ (0.5)
  \\
 $\eta K^{*-}$ & $27.3^{+9.6}_{-8.2} \pm 5.0$ & 16.2 ($\ \, 3.9$)  &
 $\eta^\prime K^{*-}$ & 14.4 (2.0)  &
 $K^{*0}\pi^-$  &  $\ \, 3.5$ ($\ \, 5.5$)  &
 $\rho^- \overline K^{0}$ &  $\ \, 8.0$ (0.4) 
  \\
 $\eta\overline K^{*0}$ & $13.8^{+5.5}_{-4.4} \pm 1.7$ & 13.2 ($\ \, 4.3$)  &
 $\eta^\prime\overline K^{*0}$ &  13.8 (1.0) & 
 $K^{*0}\pi^0$  &  $\ \, 0.6$ ($\ \, 1.3$)  &
 $\rho^0 \overline K^{0}$ &  $\ \, 6.2$ (1.1)  
  \\ 
\end{tabular}
\end{center}
\end{table}

\end{document}